\documentclass[12pt]{article}
\usepackage{amsfonts}
\usepackage{amsmath}
\usepackage{theorem}
\usepackage{enumerate}
\usepackage{latexsym}
\usepackage{graphicx}

\def\a{\alpha}
\def\b{\beta}
\def\d{\delta}
\def\c{\chi}
\def\g{\gamma}
\def\l{\lambda}

\def\e{\epsilon}

\newtheorem{teo}{Theorem}[section]

\newtheorem{tve}{Proposition}[section]

\title{\bf  Lie algebras on hyperelliptic curves
and finite-dimensional integrable systems}
\author {\sl T. Skrypnyk }

\date{ }

\begin{document}
\thispagestyle{empty}
 \maketitle
 {\it \ \ \ \ \ \ \ \ \ \ \ \ \ \ Bogoliubov Institute for Theoretical Physics,
  Kiev, Ukraine.}
\begin{center}
 {\sl e-mail:} tskrypnyk@imath.kiev.ua
\end{center}
 \begin {abstract}
We construct a new family of infinite-dimensional quasi-graded Lie
algebras on hyperelliptic curves.  We show that constructed
 algebras  possess infinite number of
invariant functions and admit a decomposition into the direct sum
of two subalgebras. These two facts together enables one to use
them to construct new integrable  finite-dimensional hamiltonian
systems. In such a way we find new integrable hamiltonian systems,
which are direct higher rank generalizations of the integrable
systems of Steklov-Liapunov, associated with the e(3) algebra and
Steklov-Veselov associated with the so(4) algebra.
\end{abstract}
\section{Introduction}
The main purpose of the present paper is to introduce  new
integrable systems of Euler-Arnold type on the finite-dimensional
Lie algebras. It is known, that  such systems admit Lax-pair
representations:
\begin{equation}\label{lp}
\dfrac{dL(t)}{dt}=[L(t),M(t)],
\end{equation}
where $L$ and $M$ are some matrix depending on the dynamical
variables  \cite{Lax}.

 Almost in all cases Lax operator depend also on
additional parameter --- so called "spectral parameter" $w$:
$L=L(w)$. This dependence permits to construct large number of
integrals of motion via the expansion of functions $I^k(w)=Tr
L(w)^k$ in the  series of the powers of the spectral parameter.
Usually dependence of Lax operator on $w$ is rational or elliptic.

The natural approach to construction of new types of integrable
systems is search for the solutions of Lax equations with other
(more complicated) dependence on the spectral parameter.
 We will
solve this problem for the case of $L(w)$ with the hyperelliptic
dependence on the spectral parameter.

 Our approach is based on the usage of infinite-dimensional Lie algebras.
It is known \cite{RST1},\cite{RST2},\cite{RST3} that group
theoretical explanation of the integrability of Lax equations on
finite-dimensional Lie algebras with the rational spectral
parameters is based on the loop algebras, i.e. algebras  of the
form $\mathfrak{g}\otimes P(\l,\l^{-1})$. In the works
\cite{Hol1},\cite{Hol2}, \cite{Hol3} it was shown, that Lax
equation with the
 elliptic spectral parameters on the algebra $so(3)$ and some its
 extensions could be obtained from the  infinite-dimensional
 Lie algebras of the special elliptic matrix-valued functions with
 the values in $so(3)$ via Kostant-Adler scheme \cite{Kos}- \cite{Adl}.

We generalize construction of \cite{Hol2} on the case of classical
matrix algebras of higher rank and investigate corresponding
finite-dimensional integrable systems. Growth of the rank of
algebra requires automatic growth of the genus of the curve. As a
result we obtain algebras of $gl(n)$- , $so(n)$- and $sp(n)$-
valued functions on the hyperelliptic curves of genus $g$, where
$n=2g+2$ or $n=2g+1$, or, to be more precise, on its double
covering. Obtained algebras have many nice properties. They are
quasi-graded. They possess central extensions. They possess
infinite number of invariant functions.
 In the rational degeneration of
the curve they coincides with the ordinary loop algebras. So they
could be viewed as "hyperelliptic '' generalizations of the loop
algebras.

There exists other higher genus generalization of the loop
algebras
--- so called Krichiver-Novikov algebras \cite{KrN} of the
matrix-valued holomorphic  functions on the Riemanian surfaces
with two punctured points. But Krichiver-Novikov algebras do not
admit Kostant-Adler scheme \cite{ScS}. Hence this generalization
can not be used for producing new integrable systems with the
spectral parameter on the higher genus curves.

Contrary to the Krichiver-Novikov algebras, our algebras admit
Kostant-Adler scheme, and hence, could be used for constructing
integrable systems.

 Using our algebra as  hidden symmetry algebra we obtain
new integrable systems  on $\mathfrak{g}\oplus\mathfrak{g}$ that
describes two interacting generalized rigid bodies. This system is
direct higher rank generalization of integrable case of Steklov
and Veselov \cite{Ves} on $so(4)=so(3)\oplus so(3)$. We also
obtain new integrable systems on the semidirect sum
$\mathfrak{g}+\mathfrak{g}$, that  generalize integrable case of
Steklov-Ljapunov on  $e(3)$ \cite{Ste}.

 Structure of the
algebra provides a possibility  to give two- dimensional
generalization of constructed integrable systems via finite-gap
extension method \cite{Hol1}-\cite{Hol3}. We will return to this
problem in our subsequent publications.

\section{Quasi-graded algebras on hyperelliptic curves}
\subsection{Construction}
\paragraph{Hyperelliptic curve embedded in $\mathbb{C}^n$.}Let us consider in the space $\mathbb{C}^n$ with the
coordinates $w_1,w_2,...., w_n$ the following system of quadrics:
\begin{equation}\label{quad}
w^2_i-w^2_j=a_j -a_i, \ \ i,j=1,n,
 \end{equation}
where $a_i$ are arbitrary complex numbers. Rank of this system is
$n-1$, so substitution: $$w^2_i=w -a_i,$$ solves these equations.
Moreover if we put $y=\prod\limits_{i=1}^n w_i$ we obtain the
equation of the hyperelliptic  curve $\mathcal{H}$:
\begin{equation}
y^2=\prod\limits_{i=1}^n(w-a_i).
 \end{equation}
Hence equations $\ref{quad}$ define embedding of the hyperelliptic
curve $\mathcal{H}$ in the linear space $C^n$. Variable $w$ is a
local parameter on the curve and $a_i$  its branching points.\\

\noindent{\it Example.}\ In the $n=3$ case all these objects have
well-known analytical description. Indeed in this case curve under
the consideration is elliptic. Its uniformization is made by the
Weierstrass $\mathfrak{p}$- function and its derivative:
$w=\mathfrak{p}(u)$,\ \  $y=1/2\mathfrak{p}'(u)$. Functions $w_i$
are  expressed via Jacobi elliptic functions \cite{CG}:
  $w_1=\dfrac{1}{sn(u)} $, \ $w_2=\dfrac{dn(u)}{sn(u)} $, \
$w_3=\dfrac{cn(u)}{sn(u)}$.

\paragraph{Classical Lie algebras.}Let $\mathfrak{g}$ denotes one of the classical matrix Lie
algebras $gl(n)$, $so(n)$ and $sp(n)$ over the field of the
complex numbers. For the subsequent we will need special form of
their bases. Let us explicitly construct them. Let $I_{i,j}\in
Mat(n,C)$ be a matrix defined as follows:
$$(I_{ij})_{ab}=\d_{ia}\d_{jb}.$$ Evidently, a basis in the
algebra $gl(n)$ could be built from the matrices $X_{ij}\equiv
I_{ij}$, $i,j\in 1,...,n$. The commutation relations in $gl(n)$
will have the following standard form:
$$[X_{i,j},X_{k,l}]=\d_{k,j} X_{i,l} -\d_{i,l} X_{k,j}.$$ The
basis in the algebra $so(n)$ could be chosen as: $X_{ij}\equiv
I_{ij}-I_{i,j}$, $i,j\in 1,...,n$, with the following commutation
relations: $$[X_{i,j},X_{k,l}]=\d_{k,j} X_{i,l} -\d_{i,l}
X_{k,j}+\d_{j,l}X_{k,i }-\d_{k,i}X_{j,l },$$ and "skew-symmetry''
property  $X_{ij}=-X_{ji}$.\\
 The basis in the algebra $sp(n)$ consists of the matrices
 $X_{ij}=I_{ij}-\e_i\e_j I_{-i,-j}$, $i,j \in -n,...,-1,1,...,n$,
 with the commutation relations:
 $$[X_{i,j},X_{k,l}]=\d_{k,j} X_{i,l}
-\d_{i,l} X_{k,j}+\e_i \e_j(\d_{j,-l}X_{k,-i}-\d_{k,-i}X_{-j,l
}),$$ and additional property $X_{i,j}= - \e_i \e_j X_{-j,-i }$,
 where $\e_j=\mathrm{sign}j$.
\paragraph {Algebras on the curve.} For the basic elements $X_{ij}$
of all three algebras $gl(n)$, $so(n)$ and $sp(n)$ we introduce
the  following algebra-valued functions on the curve
$\mathcal{H}$, or to be more precise on its double covering:
\begin{equation}\label{gen}
X^+_{ij}(w)=X_{ij}\otimes w_iw_j,\ \ \ X^-_{ij}(w)=X_{ij}\otimes
w^{-1} w_iw_j.
\end{equation}
Here we put $w_{-i}\equiv w_i$ in the case of $sp(n)$. We will
need the following definition \cite{KrN}:
\\
{\it Definition.} Infinite-dimensional Lie algebra
$\widetilde{\mathfrak{g}}$ is called $\mathbb{Z}$-quasi graded, if
there exist such $p,q\in \mathbb{Z}_{+}$  that:
\begin{equation}\label{def}
\widetilde{\mathfrak{g}}=\sum\limits_{j\in \mathbb{Z}}\widetilde{
\mathfrak{ g}}_{j},\ \ \ \ \text{such that}\ \ \ \ \
[\widetilde{\mathfrak{ g}_{i}},\widetilde{\mathfrak{
g}}_{j}]\subset \sum\limits_{k=-p}^q \widetilde{\mathfrak{
g}}_{i+j +k}.
\end{equation}

\noindent The following theorem holds true:
\begin{teo}
(i)Functions $X^{+}_{ij}(w)$ and $X^{-}_{ij}(w)$ generate
infinite-dimensional $\mathbb{Z}$ quasi-graded Lie algebra
$\widetilde{\mathfrak{g}}_{\mathcal{H}}$.\\ (ii)There exists a
decomposition: $\widetilde{\mathfrak{g}}_{\mathcal{H}}=
\widetilde{\mathfrak{g}}_{\mathcal{H}}^{+}+
\widetilde{\mathfrak{g}}_{\mathcal{H}}^{-}$, where are
$\widetilde{\mathfrak{g}}_{\mathcal{H}}^{+}$ and
$\widetilde{\mathfrak{g}}_{\mathcal{H}}^{-}$ are generated by
$X^{+}_{ij}(w)$ and $X^{-}_{ij}(w)$ correspondingly.
\end{teo}
{\bf Proof.} Let us introduce the following algebra-valued
functions on the double covering of the curve $\mathcal{H}$:
$$X^n_{ij}=X_{ij}\otimes w^{n} w_iw_j.$$ To prove the theorem we
will need the explicit form of their commutation relations:
\begin{tve}\label{com}
Elements $X^n_{ij}$ form closed algebra with the following
commutation relations:
\begin{multline}
1)\ [X_{ij}^n,X_{ij}^m]=\d_{kj}X_{il}^{n+m+1} -
\d_{il}X_{kj}^{n+m+1} +a_i \d_{il}X_{kj}^{n+m}- a_j
\d_{kj}X_{il}^{n+m}\  \ \ \text{for the  } gl(n)
\end{multline}
\begin{multline}
2)\ [X_{ij}^n,X_{ij}^m]=\d_{kj}X_{il}^{n+m+1} -
\d_{il}X_{kj}^{n+m+1} +\d_{jl}X_{ki}^{n+m+1} -
\d_{ik}X_{jl}^{n+m+1}\\
 +a_i\d_{il}X_{kj}^{n+m}- a_j \d_{kj}X_{il}^{n+m}
+a_i \d_{ik}X_{jl}^{n+m}- a_j \d_{jl}X_{ki}^{n+m}
  \ \ \text{for the } so(n)
\end{multline}
\begin{multline}
3)\ [X_{ij}^n,X_{ij}^m]=\d_{kj}X_{il}^{n+m+1} -
\d_{il}X_{kj}^{n+m+1} +\e_i\e_j(\d_{j-l}X_{k-i}^{n+m+1} -
\d_{i-k}X_{j-l}^{n+m+1})\\ +a_i\d_{il}X_{kj}^{n+m}- a_j
\d_{kj}X_{il}^{n+m} +a_i \e_i\e_j(a_i\d_{i-k}X_{j-l}^{n+m}- a_j
\d_{j-l}X_{k-i}^{n+m})
  \ \ \text{for the  } sp(n).
\end{multline}
\end{tve}
\noindent Let us  put
$$\widetilde{\mathfrak{g}}_{\mathcal{H}}=\mathrm{Span}_\mathbb{C}\{X^m_{ij},
m\in \mathbb{Z}\}.$$
 From the Proposition \ref{com}
follows that $\widetilde{\mathfrak{g}}_{\mathcal{H}}$ is
quasi-graded. Besides,  let us define the following subspaces:
$$\widetilde{\mathfrak{g}}_{\mathcal{H}}^{+}=\mathrm{Span}_\mathbb{C}\{X^m_{ij},
m\in \mathbb{Z}_+ \cup{0}\}, \ \
\widetilde{\mathfrak{g}}_{\mathcal{H}}^{-}=
\mathrm{Span}_\mathbb{C}\{X^{-m}_{ij}, m\in \mathbb{Z}_+ \}.$$
From the Proposition \ref{com} follows, that they are subalgebras.
 Taking into account that
$X^{+}_{ij}(w)=X^{0}_{ij}$, $X^{-}_{ij}(w)=X^{-1}_{ij}$ and the
Proposition \ref{com}, it is easy to see that elements
$X^{+}_{ij}(w)$ generate subalgebra
$\widetilde{\mathfrak{g}}_{\mathcal{H}}^{+}$ and elements
$X^{-}_{ij}(w)$ generate subalgebra
$\widetilde{\mathfrak{g}}_{\mathcal{H}}^{-}$. Hence
$X^{+}_{ij}(w)$ and $X^{-}_{ij}(w)$ generate
$\widetilde{\mathfrak{g}}_{\mathcal{H}}$. Theorem is proved.\\

\noindent {\it Example.}\ \ \ Let $\mathfrak{g}=so(3)$. In this
case constructed algebra will coincide with the "even'' subalgebra
of the algebra of hidden symmetry of Landau- Lifschits equations
\cite{Hol3}. Indeed, putting $X_k\equiv \e_{ijk}X_{ij}$, we obtain
the following  commutation relations: $$[X_i^n,X_j^m]=\e_{ijk}
X_{k}^{n+m+1}
 +\e_{ijk}a_k X_{k}^{n+m}.$$

 \noindent{\it Remark.} From the Proposition \ref{com}
follows, that in the rational degeneration of  $\mathcal{H}$:
$$y^2=w^n,$$
 i.e when $a_i=0$, we obtain, that $w_i=\sqrt{w}$ and, hence,
  $\widetilde{\mathfrak{g}}_{\mathcal{H}}=\widetilde{\mathfrak{g}},$
where $\widetilde{\mathfrak{g}}$ is ordinary loop algebra.
\subsection{Central extension}
To have the full analogy with the loop algebra, and keeping in
mind possible application to quantum integrable systems, in this
subsection we define central extensions of the algebras
$\widetilde{\mathfrak{g}}_{\mathcal{H}}$. First we remind the
following definition \cite{KrN}:
\\
{\it Definition.} Cocycle $\c$ on the quasigraded algebra is
called local if $$\c(X^m_{ij},X^n_{kl})=0,\ \text{ for all }\
|n+m|>K\ \text{ and for some }\ \ K\in \mathbb{Z}_+.$$

Let $(X|Y)=c_n tr XY$, where $c_n=2n\  \text{for}\ gl(n)$,
$c_n=(n-2)\ \text{for}\ so(n)$, $c_n=(2n+2)\ \text{for}\ sp(n)$,
be a standard invariant form on the classical Lie algebras
$gl(n)$, $so(n)$ or $sp(n)$. Let us define on
$\widetilde{\mathfrak{g}}_{\mathcal{H}}$ following bilinear form:
\begin{equation}\label{coc}
\c(X(w),Y(w)=\oint\limits_{\g} (X(w)|\frac{dY(w)}{dw})dw,
\end{equation}
where $1$-cycle $\g$ in the complex plane of variable $w$
encircles point $w=0$.

 Although the above bilinear form
 is standard in the theory of loop algebras, it is not
straightforward fact that it determines the cocycle on
$\widetilde{\mathfrak{g}}_{\mathcal{H}}$ due to the fact, that
 $w_i$ are not one-valued functions of  $w$.
 \noindent We prove this in the
following theorem:
\begin{teo}
(i)Bilinear form $\c$ is skew-symmetric and satisfies the
properties of cocycle, and hence, determines a central extension
of $\widetilde{\mathfrak{g}}_{\mathcal{H}}$:
$\widehat{\mathfrak{g}}_{\mathcal{H}}=\widetilde{\mathfrak{g}}_{\mathcal{H}}
+ \mathbb{C}c$ by the following formula: $$[X(w)+\a c ,Y(w)+\b
c]'=[X(w),Y(w)]+ \c(X(w),Y(w)) c.$$ (ii) Cocycle $\c $ is local
and its values on the bases elements are calculated by the
following formulas:
\begin{equation}\label{coc1}
\c(X_{ij}^m,X_{kl}^n)=\c^{mn}(a_i,a_j)(X_{ij}|X_{kl}),\ \text{
where} \end{equation}
\begin{equation}\label{coc2}
 \c^{mn}(a_i,a_j)=(n+1)\d_{n+m+2,0} -(n+1/2)(a_i
+a_j)\d_{n+m+1,0} + n a_ia_j\d_{n+m,0}. \end{equation}
\end{teo}

\noindent{\it Example.} In the case of rational degeneration
$a_i=0$ we obtain the following cocycle on the loop algebra:
$$\c(X^m_{ij},X^n_{kl})=(n+1)\d_{n+m+2,0}(X_{ij}|X_{kl}).$$ It
pass to the standard one after renaming the indices:
$m+1\rightarrow m$, $n+1\rightarrow n$.
\subsection{Coadjoint representation}
 To define the coadjoint representation we have to
define $\widetilde{\mathfrak{g}}_{\mathcal{H}}^*$. We assume, that
$\widetilde{\mathfrak{g}}_{\mathcal{H}}^*\subset
\mathfrak{g}\otimes A$, where $A$ is an algebra of function on the
double covering of the curve $\mathcal{H}$.
 Let us define invariant pairing between
$L(w)\in\widetilde{\mathfrak{g}}_{\mathcal{H}}^*$ and $X(w) \in
\widetilde{\mathfrak{g}}_{\mathcal{H}}$ in the following way:
\begin{equation}\label{pair}
\langle X(w),L(w)\rangle_f=c_n
res_{w=0}f^{-1}(w)y^{-1}(w)(X(w)|L(w)),
\end{equation}
where $f(w)$ is arbitrary function on the curve $\mathcal{H}$.
 It is easy to show, that element dual to $X_{ij}^{-m}$ with respect
 to this pairing is $Y_{ij}^{m}\equiv (X_{ij}^{-m})^*=\dfrac{w^{m-1}f(w) y(w)}{w_iw_j}X_{ij}^*$. Hence
 the general element of the dual space has the following form:
\begin{equation}
L(w)=\sum\limits_{m\in Z} \sum\limits_{i,j=1}^{n}l^m_{ij}
\frac{w^{m-1}f(w) y(w)}{w_iw_j}X_{ij}^*
\end{equation}
Coadjoint action of algebra
$\widetilde{\mathfrak{g}}_{\mathcal{H}}$ on its dual space
$\widetilde{\mathfrak{g}}_{\mathcal{H}}^*$  coincides with
commutator:
\begin{equation}\label{coad}
ad^*_{X(w)} L(w)=[L(w),X(w)].
\end{equation}
 Explicitly coadjoint action of algebra
 has the following form:
\begin{multline*}
1)\ [X_{ij}^n,Y_{kl}^{m}]=\d_{kj}Y_{il}^{n+m+1} -
\d_{il}Y_{kj}^{n+m+1} +a_j \d_{il}Y_{kj}^{n+m}- a_i
\d_{kj}Y_{il}^{n+m}\  \ \ \text{for the  } gl(n)
\end{multline*}
\begin{multline*}
2)\ [X_{ij}^n,Y_{kl}^{m}]=\d_{kj}Y_{il}^{n+m+1} -
\d_{il}Y_{kj}^{n+m+1} +\d_{jl}Y_{ki}^{n+m+1} -
\d_{ik}Y_{jl}^{n+m+1}\\
 +a_j\d_{il}Y_{kj}^{n+m}- a_i \d_{kj}Y_{il}^{n+m}
+a_j \d_{ik}Y_{jl}^{n+m}- a_i \d_{jl}Y_{ki}^{n+m}
  \ \ \text{for the } so(n)
\end{multline*}
\begin{multline*}
3)\ [X_{ij}^n,Y_{kl}^{m}]=\d_{kj}Y_{il}^{n+m+1} -
\d_{il}Y_{kj}^{n+m+1} +\e_i\e_j(\d_{j-l}Y_{k-i}^{n+m+1} -
\d_{i-k}Y_{j-l}^{n+m+1})\\ +a_j\d_{il}Y_{kj}^{n+m}- a_i
\d_{kj}Y_{il}^{n+m} + \e_i\e_j(a_j\d_{i-k}Y_{j-l}^{n+m}- a_i
\d_{j-l}Y_{k-i}^{n+m})
  \ \ \text{for the  } sp(n).
\end{multline*}
\\
 It is evident from the above formulas, that
 $\widetilde{\mathfrak{g}}_{\mathcal{H}}^*$ is
 a quasi-graded $\widetilde{\mathfrak{g}}_{\mathcal{H}}$-module:
$$\widetilde{\mathfrak{g}}_{\mathcal{H}}^*=\sum\limits_{m\in
\mathbb{Z}}(\widetilde{\mathfrak{g}}_{\mathcal{H}}^*)_m,$$ where
$(\widetilde{\mathfrak{g}}_{\mathcal{H}}^*)_m=Span_C\{Y^m_{ij}|i,j=1,...,n\}.$
\\
{\it Remark 1.} Putting $f(w)= y(w)$ in the definition of the
elements of dual space we obtain, that
$\widetilde{\mathfrak{g}}_{\mathcal{H}}^*\subseteq
\widetilde{\mathfrak{g}}_{\mathcal{H}}$. If $f(w)\neq y(w)$, then
generally speaking, spaces
$\widetilde{\mathfrak{g}}_{\mathcal{H}}^*$ and
$\widetilde{\mathfrak{g}}_{\mathcal{H}}$, viewing as subspaces in
the algebra of functions on the double covering of the curve
$\mathcal{H}$,  do not coincide.
\\
{\it Remark 2.} If $f=1$ and $\mathfrak{g}=so(n)$ elements of
$\widetilde{\mathfrak{g}}_{\mathcal{H}}
+\widetilde{\mathfrak{g}}_{\mathcal{H}}^*$ form a closed
algebra.This will be analogue of the algebra of  hidden symmetry
of Landau- Lifschits  equations \cite{Hol1}. Unfortunately, for
$n>4$ it does not admit a Kostant-Adler scheme and can not be used
for the construction of integrable systems. That is why we will
not consider it here.
\\ From the explicit form of coadjoint action (\ref{coad}) follows
the next statement:
\begin{tve}\label{cas}
Functions $I^k_m (L(w))= res_{w=0} w^{-m-1}Tr L(w)^k$, where $m\in
\mathbb{Z}$, are invariants of coadjoint representation.
\end{tve}
Hence constructed Lie algebras not only admit decomposition into
the direct sum of two subalgebras but also possess infinite number
of invariant functions. This permits to use them in  construction
of integrable systems.
\section{Integrable systems from hyperelliptic algebras}
\subsection{Poisson structures and Poisson subspaces.}
\paragraph{First Lie-Poisson structure.}In the space $\widetilde{\mathfrak{g}}_{\mathcal{H}}^*$ it is
possible to define many Lie-Poisson structures using different
pairings. We will use the  pairing (\ref{pair}) with $f(w)=w$:
\begin{equation}\label{pair1}
\langle X(w),L(w)\rangle_{-1}=c_n
res_{w=0}w^{-1}y^{-1}(w)(X(w)|L(w)).
\end{equation}
It defines  brackets on
$P(\widetilde{\mathfrak{g}}_{\mathcal{H}}^*)$ in the following
  way:
\begin{equation}\label{pair1}
\{F(L),G(L)\}= \sum\limits_{l,m\in
Z}\sum\limits_{i,j,p,s=1}^n\langle
L(w),[X_{ij}^{-l},X_{ps}^{-m}]\rangle_{-1}\dfrac{\partial G
}{\partial l_{ij}^l }\dfrac{\partial F }{\partial l_{ps}^m }
\end{equation}
The following Corollary of the Proposition \ref{cas} holds true:
\begin{tve}
Functions $I^k_m (L(w))$ are central for  brackets  $\{ \ ,\ \}$.
\end{tve}
Taking into account that $l_{ij}^m=\langle
L(w),X^{-m}_{ij}\rangle_{-1}$, it is easy to show, that for the
coordinate functions $l_{ij}^m$ these brackets will have the
following form:
\begin{multline}
1)\ \{l_{ij}^n,l_{ij}^m\}=\d_{kj}l_{il}^{n+m-1} -
\d_{il}l_{kj}^{n+m-1} +a_i \d_{il}l_{kj}^{n+m}- a_j
\d_{kj}l_{il}^{n+m}\  \ \ \text{for the  } gl(n)
\end{multline}
\begin{multline}
2)\ \{l_{ij}^n,l_{ij}^m\}=\d_{kj}l_{il}^{n+m-1} -
\d_{il}l_{kj}^{n+m-1} +\d_{jl}l_{ki}^{n+m-1} -
\d_{ik}l_{jl}^{n+m-1}\\
 +a_i\d_{il}l_{kj}^{n+m}- a_j \d_{kj}l_{il}^{n+m}
+a_i \d_{ik}l_{jl}^{n+m}- a_j \d_{jl}l_{ki}^{n+m}
  \ \ \text{for the } so(n)
\end{multline}
\begin{multline}
3)\ \{l_{ij}^n,l_{ij}^m\}=\d_{kj}l_{il}^{n+m-1} -
\d_{il}l_{kj}^{n+m-1} +\e_i\e_j(\d_{j-l}l_{k-i}^{n+m-1} -
\d_{i-k}l_{j-l}^{n+m-1})\\ +a_i\d_{il}l_{kj}^{n+m}- a_j
\d_{kj}l_{il}^{n+m} + \e_i\e_j(a_i\d_{i-k}l_{j-l}^{n+m}- a_j
\d_{j-l}l_{k-i}^{n+m})
  \ \ \text{for the  } sp(n).
\end{multline}
From the explicit form of the Poisson brackets  follows, that
Lie-Poisson brackets in the subspaces
$(\widetilde{\mathfrak{g}}_{\mathcal{H}}^*)_{-}=
\sum\limits_{m=0}^{\infty}
(\widetilde{\mathfrak{g}}_{\mathcal{H}}^*)_{-m}$  and
$(\widetilde{\mathfrak{g}}_{\mathcal{H}}^*)_{+}=
\sum\limits_{m=1}^{\infty}
(\widetilde{\mathfrak{g}}_{\mathcal{H}}^*)_{m}$, up to the reverse
of the sign of the upper indices, repeat commutation relations of
the algebras $\widetilde{\mathfrak{g}}_{\mathcal{H}}^{+}$ and
$\widetilde{\mathfrak{g}}_{\mathcal{H}}^{-}$ correspondingly.
This, of course, is the result of the following duality:
$(\widetilde{\mathfrak{g}}_{\mathcal{H}}^{+})^*=(\widetilde{\mathfrak{g}}_{\mathcal{H}}^*)_{-}$,
$(\widetilde{\mathfrak{g}}_{\mathcal{H}}^{-})^*=(\widetilde{\mathfrak{g}}_{\mathcal{H}}^*)_{+}$.
\paragraph{Second Lie-Poisson structure.} Let us introduce into the space
$\widetilde{\mathfrak{g}}_{\mathcal{H}}^{*}$ a new Poisson
brackets  $\{\ ,\ \}_0$, which are a Lie-Poisson brackets for the
algebra $\widetilde{\mathfrak{g}}_{\mathcal{H}}^0$, where
$\widetilde{\mathfrak{g}}_{\mathcal{H}}^0=
\widetilde{\mathfrak{g}}_{\mathcal{H}}^- \ominus
\widetilde{\mathfrak{g}}_{\mathcal{H}}^+$. Explicitly, this
brackets have the following form:
\begin{gather*}
 \{l^n_{ij},l^{m}_{kl}\}_0=
-\{l^n_{ij},l^{m}_{kl}\},\ n,m > 0 ,\ \ \
 \{l^n_{ij},l^{m}_{kl}\}_0=
\{l^n_{ij},l^{m}_{kl}\},\ n,m \leq 0,\\
 \{l^n_{ij},l^{m}_{kl}\}_0=0
, m\leq 0,n > 0\  \text{or}\
 n \leq 0,m  > 0\ .
\end{gather*}
 Let subspace
$\mathcal{M}_{s,p}\subset\widetilde{\mathfrak{g}}_{\mathcal{H}}^{*}$
be defined as follows: $$
\mathcal{M}_{s,p}=\sum\limits_{m=-s+1}^{p}
(\widetilde{\mathfrak{g}}_{\mathcal{H}}^*)_{m}.$$ Brackets $\{\ ,\
\}_0$ could be correctly restricted to $\mathcal{M}_{s,p}$. It
follows from the next Proposition:
\begin{tve}\label{idea}
Subspaces $\mathcal{J}_{p,s}=\sum\limits_{m=-\infty}^{-p-1}
(\widetilde{\mathfrak{g}}_{\mathcal{H}})_{m}+
\sum\limits_{m=s}^{\infty}
(\widetilde{\mathfrak{g}}_{\mathcal{H}})_{m}$ are ideals in
$\widetilde{\mathfrak{g}}_{\mathcal{H}}^0$.
\end{tve}
{\bf Proof.} It follows from the explicit form of commutation
relations in $\widetilde{\mathfrak{g}}_{\mathcal{H}}^0$.
\\

Now we are ready to prove the following important theorem:
\begin{teo}
Functions $\{I^k_m(L)\}$ commutes with respect to the restriction
of the brackets $\{\ ,\ \}_0$ on $\mathcal{M}_{s,p}$.
\end{teo}
{\bf Proof.} It follows from the combination of Kostant-Adler
scheme and previous  Proposition. Indeed, due to the fact, that
$\{I^k_m(L)\}$ are Casimir functions on
$\widetilde{\mathfrak{g}}_{\mathcal{H}}^*$ they form a commutative
subalgebra with respect to the brackets $\{\ ,\ \}_0$ \cite{RST1}.
Hence they will stay commutative after the restriction to
$\mathcal{M}_{s,p}=(\widetilde{\mathfrak{g}}_{\mathcal{H}}^0/\mathcal{J}_{p,s})^*$,
due to the fact, that projection onto quotient algebra is a
canonical homomorphism. \\

\noindent Hamiltonian equations will have the following form:
\begin{equation}
\frac{dl^{k}_{ij}}{dt}=\{l^k_{ij},H(l^{m}_{kl})\}_0,
\end{equation}
 where
hamiltonian $H$ is one of the functions $I^k_m$ or their linear
combination. These equations could be written in the Lax form
\cite{RST3}:
\begin{equation}
\frac{dL(w)}{dt}=[L(w),M(w)],
\end{equation}
where $L(w)\in \mathcal{M}_{s,p}$, and second operator is defined
as follows: $M(w)=(P_--P_+)\nabla H(L(w))$. Here $P_{\pm}$ are
projection operators on the subalgebra
$\widetilde{\mathfrak{g}}_{\mathcal{H}}^{\pm}$,
\begin{equation}\label{sgrad}
\nabla H(L(w))=\sum\limits_{k=-p}^{s-1}\sum\limits_{ij=1}^n
\dfrac{\partial H }{\partial l_{ij}^k }X_{ij}^{-k}
\end{equation}
 is an algebra-valued gradient of $H$.
\section{ Integrable systems in finite-dimensional quotients.}
The most interesting from the physical point of view examples
usually arise in the spaces $\mathcal{M}_{s,p}$ with small $s$ and
$p$. We will consider the case $|s+p|\leq 2$.
 We will assume, that
curve $\mathcal{H}$ is nondegenerated, i.e. $a_i\neq a_j$ for
$i\neq j$. This requirement is necessary for completeness of the
family of constructed commuting functions.

The basic algebra in all examples will be $\mathfrak{g}=so(n)$,
but analogous results are valid for $\mathfrak{g}=gl(n)$ and
$\mathfrak{g}=sp(n)$. We chose for all examples $so(n)$ algebra
only because correspondent integrable systems are the most direct
generalizations of classical integrable systems connected with
$so(4)$ and $e(3)$.

\subsubsection{Generalized interacting  tops.}
 Let us  consider subspace
$\mathcal{M}_{1,1}$. In the case $a_i\neq 0$, and as it follows
from the arguments below,
 $(\widetilde{\mathfrak{g}}_{\mathcal{H}}^+/\mathcal{J}_{1,1})\simeq
 \mathfrak{g}\oplus\mathfrak{g}$, hence
$\mathcal{M}_{1,1}= (\mathfrak{g}\oplus\mathfrak{g})^*.$
Corresponding Lax operator $L(w)\in \mathcal{M}_{0,1}$ has the
following form: $$L(w)=\sum\limits_{i,j=1}^n (l^{(0)}_{ij}+ w
l^{(1)}_{ij})\dfrac{ y(w)}{w_i w_j}X^*_{ij}.$$ Let us again
consider $so(n)$ case and put $ X^*_{ij}= X_{ij}$. Lie-Poisson
brackets between the coordinate functions $l^{(1)}_{ij}$ have the
following form:  $$ \{l_{ij}^{(0)},l_{kl}^{(0)}\}=
 -a_i\d_{il}l_{kj}^{(0)}+ a_j \d_{kj}l_{il}^{(0)}
-a_i \d_{ik}l_{jl}^{(0)}+ a_j \d_{jl}l_{ki}^{(0)},$$
$$\{l^{(1)}_{i,j},l^{(1)}_{k,l}\}=\d_{k,j} l^{(1)}_{i,l} -\d_{i,l}
l^{(1)}_{k,j}+\d_{j,l}l^{(1)}_{k,i }-\d_{k,i}l^{(1)}_{j,l },$$
$$\{l_{ij}^{(0)},l_{kl}^{(1)}\}=0.$$ Putting $b_i=a_i^{1/2}$ and
making the change of variables: $l_{ij}=l^{(1)}_{i,j}$,
$m_{ij}=\dfrac{l_{ij}^{(0)}}{b_i b_j}$, we obtain canonical
coordinates of the direct sum of two algebras $so(n)$:
$$\{m_{i,j},m_{k,l}\}=\d_{k,j} m_{i,l} -\d_{i,l}
m_{k,j}+\d_{j,l}m_{k,i }-\d_{k,i}m_{j,l },$$
$$\{l_{i,j},l_{k,l}\}=\d_{k,j} l_{i,l} -\d_{i,l}
l_{k,j}+\d_{j,l}l_{k,i }-\d_{k,i}l_{j,l },$$
$$\{l_{ij},m_{kl}\}=0.$$
 \noindent Commuting integrals are
constructed using expansions in the powers of $w$ of the
functions: $I_k(w)=Tr(L(w))^k$. As in the previous example, we are
mainly  interested in the quadratic integrals. Let $$h(w)\equiv
I_2(w)=\sum\limits_{s=0}^{n} h_{s}(l^{(1)}_{ij})w^s
=\sum\limits_{ij}(\prod_{k\neq
i,j}(w-a_k))(l^{(0)}_{ij}+wl^{(1)}_{ij})^2.$$
 By straightforward calculations we obtain:\\

$h_0=(-1)^{n-2}\sum\limits_{i,j=1}^n\dfrac{a_1 a_2 ...\  a_n}{a_i
a_j}(l^{(0)}_{ij})^2$\\

$h_1=(-1)^{n-1}\sum\limits_{i,j=1}^n(\sum\limits_{k\neq
i,j}\dfrac{a_1 a_2...\ a_n}{a_k})\dfrac{(l^{(0)}_{ij})^2}{a_i
a_j}-2 \dfrac{a_1 a_2 ...\  a_n}{a_i
a_j}l^{(0)}_{ij}l^{(1)}_{ij}$\\

 $\cdots \cdots  \cdots  \cdots  \cdots \cdots$\\

 $h_{n-1}=- \sum\limits_{i,j=1}^n(\sum\limits_{k=1}^n
a_k-(a_i+a_j))(l^{(1)}_{ij})^2-2l^{(0)}_{ij}l^{(1)}_{ij}$\\

 $h_{n}=
\sum\limits_{i,j=1}^n(l^{(1)}_{ij})^2$. \\ Making the described
above replacement of variables we will have:\\

$h_0=(-1)^{n-2}(b^2_1 b^2_2...\ b^2_n)\sum\limits_{i,j=1}^n
m_{ij}^2$\\

$h_1=(-1)^{n-1}\sum\limits_{i,j=1}^n(\sum\limits_{k\neq
i,j}\dfrac{b^2_1 b^2_2...\ b^2_n}{b^2_k})(m_{ij})^2-2 \dfrac{b^2_1
b^2_2...\ b^2_n}{b_i b_j}m_{ij}l_{ij}$\\

 $\cdots \cdots  \cdots  \cdots  \cdots \cdots$\\

 $h_{n-1}= -\sum\limits_{i,j=1}^n(\sum\limits_{k=1}^n
b^2_k-(b^2_i+b^2_j))l_{ij}^2-2b_ib_jm_{ij}l_{ij}$\\

 $h_{n}=
\sum\limits_{i,j=1}^n(l_{ij})^2$. \\ It is evident, that functions
$h_0$ and $h_n$ are invariants. Functions $h_1$,...,$h_{n-1}$
generate non-trivial hamiltonian flows. For the hamiltonian of the
generalized interacting rigid bodies we can take either $h_{n-1}$
or $h_1$. Correspondent $M$-operator and Lax equations are
calculated straightforwardly. \noindent\paragraph{ Steklov-Veselov
($n=3$) case.} In this case, making standard of variables
$l_i=\e_{ijk}l_{ij}$, $m_i=\e_{ijk}m_{ij}$ we obtain the following
set  of commuting functions:

 $h_0= \sum\limits_{k=1}^n
m_{k}^2$, \ \  $h_3= \sum\limits_{k=1}^n l_{k}^2$,

$h_1=(\sum\limits_{k=1}^n(\dfrac{b_1 b_2b_3}{b^2_k})(m_k)^2-2 b_k
m_k l_k),$

 $h_{2}= \sum\limits_{i,j=1}^n(
b^2_kl_{k}^2-2\dfrac{b_1b_2 b_3}{b_k}m_{k}l_{k}).$

\noindent Here $h_0$, $h_3$ --- invariant functions and
$h_1$,$h_2$ are two independent integrals discovered by Veselov
\cite{Ves}.
 Commutation relations between coordinates $l_k$ and $m_k$ of $so(4)$
are standard:

 $\{l_i,l_j\}=\e_{ijk} l_k$, \  $\{m_i,m_j\} =\e_{ijk} m_k$, \
$\{m_i,l_j\} =0$.

\subsubsection{Generalized Steklov-Ljapunov case.}
The last class of  integrable systems we wish to consider here
will be integrable systems that generalize classical integrable
system of Steklov-Ljapunov on $e(3)=so(3)+R^3$.
 Let us  consider subspace
$\mathcal{M}_{0,2}=
 (\widetilde{\mathfrak{g}}_{\mathcal{H}}^+/\mathcal{J}_{2,0})^*$.
 It is easy to show that $\mathcal{M}_{0,2}= (\mathfrak{g}+\mathfrak{g})^*.$ Corresponding
Lax operator $L(w)\in \mathcal{M}_{0,2}$ has the following form:
$$L(w)=w(\sum\limits_{i,j=1}^n (l^{(1)}_{ij}+ w
l^{(2)}_{ij})\dfrac{ y(w)}{w_i w_j}X^*_{ij}).$$ We will again be
concentrated on $\mathfrak{g}=so(n)$ case and put $ X^*_{ij}=
X_{ij}$. Lie-Poisson brackets between coordinate functions are
following: $$ \{l_{ij}^{(1)},l_{ij}^{(1)}\}=\d_{kj}l_{il}^{(1)} -
\d_{il}l_{kj}^{(1)} +\d_{jl}l_{ki}^{(1)} - \d_{ik}l_{jl}^{(1)}\\
 +a_i\d_{il}l_{kj}^{(2)}- a_j \d_{kj}l_{il}^{(2)}
+a_i \d_{ik}l_{jl}^{(2)}- a_j \d_{jl}l_{ki}^{(2)},$$
 $$\{l_{ij}^{(1)},l_{ij}^{(2)}\}=\d_{kj}l_{il}^{(2)} -
\d_{il}l_{kj}^{(2)} +\d_{jl}l_{ki}^{(2)} - \d_{ik}l_{jl}^{(2)},\ \
\ \{l_{ij}^{(2)},l_{ij}^{(2)}\}=0.$$

 \noindent Commuting integrals are
constructed using expansion in the powers of $w$ of the functions:
$I_k(w)=Tr(L(w))^k$. We are again interested mainly in quadratic
integrals. Let $$h(w)\equiv I_2(w)=w^2\sum\limits_{s=0}^{n}
h_{s+2}(l^{(1)}_{ij})w^s = w^2 \sum\limits_{ij}(\prod_{k\neq
i,j}(w-a_k))(l^{(1)}_{ij}+wl^{(2)}_{ij})^2.$$ Direct calculations
give:\\

$h_2=(-1)^{n-2}\sum\limits_{i,j=1}^n\dfrac{a_1 a_2 ...\  a_n}{a_i
a_j}(l^{(1)}_{ij})^2$\\

 $\cdots \cdots  \cdots  \cdots  \cdots \cdots$\\

 $h_{n+1}=(-1)( \sum\limits_{i,j=1}^n(\sum\limits_{k=1}^n
a_k-(a_i+a_j))(l^{(2)}_{ij})^2-2l^{(1)}_{ij}l^{(2)}_{ij})$\\

 $h_{n+2}=
\sum\limits_{i,j=1}^n(l^{(2)}_{ij})^2$.\\
 Change of variables:
$l^{(1)}_{ij}= l_{ij}- 1/2 (a_i +a_j)p_{ij},$ $l^{(2)}_{ij}=
p_{ij}$ transform described above brackets to the canonical
brackets on the half-direct sum $so(n) +so(n)$: $$
\{l_{ij},l_{kl}\}=\d_{kj}l_{il} - \d_{il}l_{kj} +\d_{jl}l_{ki} -
\d_{ik}l_{jl},$$
 $$\{l_{ij},p_{kl}\}=\d_{kj}p_{il} -
\d_{il}p_{kj} +\d_{jl}p_{ki} - \d_{ik}p_{jl},$$
 $$\{p_{ij},p_{kl}\}=0.$$
After such transformation we obtain  the  following set of
hamiltonians:

 $h_2=(-1)^{n-2}\sum\limits_{i,j=1}^n\dfrac{a_1 a_2
...\  a_n}{a_i a_j}(l_{ij}-1/2(a_i+a_j)p_{ij})^2$\\

 $\cdots \cdots  \cdots  \cdots  \cdots \cdots$\\

 $h_{n+1}=(-1)(\sum\limits_{k=1}^n
a_k)( \sum\limits_{i,j=1}^n p_{ij}^2)-2(\sum\limits_{i,j=1}^n
l_{ij}p_{ij})$\\

 $h_{n+2}=
\sum\limits_{i,j=1}^n p^{2}_{ij}$.
\\
Last two functions are invariant functions. First $n-1$ give
non-trivial flows on $\mathfrak{g}^*$.
 We will chose function $H=h_2$ for the hamiltonian function.
Correspondent $M$ operator is:
$$M(w)=2\sum\limits_{i,j=1}^{n}\dfrac{a_1 a_2 ...\ a_n}{a_i
a_j}(l_{ij}-1/2(a_i+a_j)p_{ij})w^{-1}w_iw_j X_{ij}.$$ Lax equation
have the standard form (\ref{lp}).

\noindent\paragraph{ Steklov-Ljapunov ($n=3$) case.} In this case,
making standard replacement of variables $l_k=\e_{ijk} l_{ij}$,
$p_k=\e_{ijk} p_{ij}$  we obtain:
$$H=(-1)^{n-2}\sum\limits_{k=1}^n a_k (l_{k}-1/2(a_1+a_2+a_3- a_k
)p_{k})^2.$$

\noindent Up to the rescaling of momenta $p_i$: $p_i\rightarrow
2\sigma p_i$ hamiltonian $H$ coincides with the hamiltonian of
Steklov-Ljapunov system in the form of Kotter \cite{Kott}.

\subsection*{Acknowledgments.}Autor is grateful to P. I. Holod for
many helpful discussions. The research described in this
publication was possible in part by the Award number UP1-2115 of
the U.S. Civilian Research and Development Foundation (CRDF) for
independent states of the former Soviet Union.

\end{document}